\documentstyle[12pt]{article}
\def\be{\begin{equation}}
\def\ee{\end{equation}}
\def\ol{\overline}

\def\barr{\begin{array}}
\def\earr{\end{array}}
\def\bearr{\begin{eqnarray}}
\def\eearr{\end{eqnarray}}
\def\h{\hbox}
\def\l{\left}
\def\r{\right}
\def\non{\nonumber}
\def\wt{\widetilde}
\begin{document}
\baselineskip=7mm
\begin{titlepage}
\title{{\vskip -1in  {\hskip 5in {\large PRL-TH-95/21}}}\\
\vskip 1in
Automatic $CP$ Invariance and Flavor Symmetry}
\author{Gautam Dutta and Anjan S. Joshipura\\
Theory Group, Physical Research Laboratory\\
Navrangpura, Ahmedabad 380 009, India}
\date{}
\maketitle
\begin{abstract}
The approximate conservation of $CP$ can be naturally understood
if it arises as an automatic symmetry of the renormalizable
Lagrangian. We present a specific realistic example with this
feature. In this example, the global Peccei-Quinn symmetry and
gauge symmetries of the model make the renormalizable Lagrangian
$CP$ invariant but allow non zero hierarchical masses and mixing
among the three generations. The left-right and a horizontal
$U(1)_H$ symmetry is imposed to achieve this. The
non-renormalizable interactions invariant under these symmetries
violate $CP$ whose magnitude can be in the experimentally
required range if $U(1)_H$ is broken at very high, typically,
near the grand unification scale.
\end{abstract}
{\vskip 1in pacs\#: 11.30.Er, 11.30.Hv, 12.15.Ff, 12.60.-i, 12.60.Cn,
 12.60.Fr}
\thispagestyle{empty}
\end{titlepage}
\newpage
The $SU(3) \otimes SU(2) \otimes U(1) $ symmetry associated with the 
Standard Model (SM) of electroweak interactions is known to be inadequate 
for explaining fermionic masses
and mixing. The gauge symmetry of the SM can accommodate these masses and in
particular the $CP$ violation \cite{km} but does not provide any theoretical
understanding of mass hierarchy or of approximate $CP$ conservation.
Some understanding of these issues can be obtained by imposing additional
symmetries acting in the space of fermionic flavors. Such horizontal 
symmetries are
known \cite{wali,ross} to lead to desired patterns of fermionic masses  and
mixing.
It can also help in understanding the approximate conservation
of $CP$. The aim of this note is to discuss this aspect of horizontal symmetry
through an example. In this example, the exact conservation of a horizontal 
$U(1)$
symmetry leads automatically to a $CP$ conserving theory while its breakdown 
at very
high scale leads to the observed $CP$ violation. 

Ideally one would like to have $CP$ as an automatic symmetry of the 
renormalizable
Lagrangian in analogy with the baryon and the lepton number symmetries 
which are
consequences of the gauge structure and the field content in the standard
 model.
This actually happens in a special case with two
generations of fermions \cite{km,weinberg} . In this case, the most
general Lagrangian invariant under the SM interactions is automatically $CP$
invariant if there is only one Higgs doublet or if there are two Higgs doublets
but natural flavor conservation is imposed as an additional
 symmetry \cite{weinberg}. 
This feature however gets spoiled when
one introduces the third generation.  

In principle the presence of the third generation need not spoil the
 $CP$ invariance if Yukawa
couplings are suitably restricted. To be realistic, these restrictions
 must however
be such that all masses and mixing angles are non-zero and hierarchical
 in accordance 
with the observed
pattern. This can be accomplished if additional gauge interactions 
are postulated.

We will present
an explicit example where the same horizontal symmetry gives Fritzsch
 structure 
\cite{fritzsch} for the
quark mass matrices and also leads automatically to a $CP$ invariant
 Lagrangian. In
realistic case, one needs $CP$ violation as well as deviations from the
 Fritzsch
structure \cite{nir}. 
Both these occur through non-renormalizable interactions when the horizontal
symmetry is broken at very high scale. 
The smallness of $CP$ violation in this case is
thus intimately linked to the scale of horizontal symmetry breaking. 

Our example requires extension of $SU(3) \otimes SU(2) \otimes U(1) $ to a
left-right symmetric theory \cite{lr}. 
In addition to the 
$G_{LR} = SU(3) \otimes SU(2)_L \otimes SU(2)_R \otimes U(1)_{(B-L)} $ 
group we need to impose a horizontal symmetry $U(1)_H$
and the Peccei-Quinn \cite{pq} (PQ) symmetry $U(1)_{PQ}$ in order to get
a fully $CP$ invariant theory. 

The $U(1)_H$ is a gauged horizontal
symmetry which is chosen to obtain texture zeroes in the quark mass matrices.
The choice of $U(1)_H$ is constrained by the requirement of anomaly
cancellation. Anomalies are seen to cancel if
one chooses the $U(1)_H$ charges (1, 0, -1) for the left handed quark fields
denoted in the weak basis by $q'_{iL}$. 
The corresponding right handed fields are chosen
to have opposite $U(1)_H$ values.  We need
to introduce three bi-doublet Higgs fields $\Phi_{\alpha}$ with 
the $U(1)_H$ charges (1,
-1, -2). These Higgs fields are needed in order to obtain essentially real but
non-trivial quark mass matrices with non-vanishing masses and mixing angles. 

The $U(1)_{PQ}$ is a global Peccei Quinn symmetry which serves dual 
purpose here. It
allows rotation of the strong $CP$ violating angle
 $\theta$ \cite{pq} and it also 
forbids some
crucial couplings in the Yukawa and Higgs sectors. 
Under the PQ symmetry, $q'_{iR}
\rightarrow e^{i\beta} q'_{iR}$ and 
$\Phi_\alpha\rightarrow e^{-i\beta} \Phi_\alpha$. Rest
of the fields remain invariant. Given this choice, the most general
 $G\equiv G_{LR}
\otimes U(1)_{PQ} \otimes U(1)_H $ invariant Yukawa couplings can be written as
\be
-{\cal L}_Y = \ol{q}'_L \Gamma_\alpha \Phi_\alpha q'_R + H.C. \label{yuk}
\ee
with 
\be
\Gamma_1 = \left(\barr{ccc} 0 & a & 0\\ a^* & 0 & 0 \\ 0 & 0 & 0\\ 
\earr \right);
\Gamma_2 = \left(\barr{ccc} 0 & 0 & 0\\ 0 & 0 & b \\ 0 & b^* & 0\\ 
\earr \right);
\Gamma_3 = \left(\barr{ccc} 0 & 0 & 0\\ 0 & 0 & 0 \\ 0 & 0 & c \\ 
\earr \right);
                                                             \label{gamma}    
\ee 
We have imposed here the conventional discrete parity \cite{lr} 
$ q'_L\leftrightarrow q'_R$ and
$\Phi_\alpha \leftrightarrow \Phi_\alpha ^{\dag}$. 
$CP$ is not imposed as a symmetry and hence the couplings $a$, $b$  appearing
 in
$\Gamma_\alpha$ are complex in general. But their phases can be rotated away
leaving a $CP$ invariant Lagrangian. In order to show this, we first
 concentrate on the
$G$ invariant scalar potential for the fields $\Phi_\alpha$ and 
$\widetilde{\Phi}_\alpha = \tau_2 \Phi^* _\alpha \tau_2$ :
\bearr 
V_1(\Phi ) &=& \mu^2_{\alpha} tr(\Phi_{\alpha}^{\dag} \Phi_{\alpha}) 
               + \lambda _{\alpha}\{tr(\Phi_{\alpha}^{\dag}\Phi_{\alpha})\}^2 
                                             \non \\        
         & &  + \lambda _{1\alpha ,\beta } 
               tr(\Phi_{\alpha}^{\dag} \widetilde{\Phi}_{\beta}) 
               tr(\widetilde{\Phi}_{\alpha}^{\dag} \Phi_{\beta}) \non \\
         & & + \rho_{1\alpha} tr(\Phi_{\alpha } \Phi_{\alpha } ^{\dag } 
                                 \Phi_{\alpha } \Phi_{\alpha } ^{\dag })
              + \rho_{2\alpha} tr(\Phi_{\alpha } ^{\dag } 
\widetilde{\Phi}_{\alpha } 
                  \widetilde{\Phi}_{\alpha } ^{\dag } \Phi_{\alpha })  
             + \rho_{3\alpha} tr(\Phi_{\alpha }
 \widetilde{\Phi}_{\alpha } ^{\dag } 
              \widetilde{\Phi}_{\alpha } \Phi_{\alpha } ^{\dag }) 
                                                          \non \\
         & & + \sum_{\alpha \neq \beta} \l\{
              \lambda _{2\alpha ,\beta } tr(\Phi_{\alpha}^{\dag} \Phi_{\beta}) 
               tr(\Phi_{\beta}^{\dag} \Phi_{\alpha})   
             + \lambda _{3\alpha ,\beta }  tr(\Phi_{\alpha}^{\dag}
             \Phi_{\alpha}) 
               tr(\Phi_{\beta}^{\dag} \Phi_{\beta}) \r. \non \\
         & &  +     \delta_{1\alpha \beta} 
                  tr(\Phi_{\alpha } ^{\dag }\Phi_{\beta }
                     \Phi_{\beta } ^{\dag }\Phi_{\alpha })    
             +\delta'_{1\alpha \beta} 
                       tr(\Phi_{\beta } ^{\dag }\Phi_{\beta }
                          \Phi_{\alpha } ^{\dag }\Phi_{\alpha }) \non \\ 
         & & + \delta_{2\alpha \beta} 
                       tr(\Phi_{\alpha } ^{\dag }\widetilde{\Phi}_{\beta }
                          \widetilde{\Phi}_{\alpha } ^{\dag }\Phi_{\beta })
             +\delta'_{2\alpha \beta} 
                       tr(\wt{\Phi}_{\alpha } ^{\dag }\wt{\Phi}_{\beta }
                          \Phi_{\alpha } ^{\dag }\Phi_{\beta }) \non \\
          & & + \l. \delta_{3\alpha \beta} 
                       tr(\Phi_{\alpha } \wt\Phi_{\beta } ^{\dag}
                          \wt{\Phi}_{\beta } \Phi_{\alpha } ^{\dag})
              +\delta'_{3\alpha \beta} 
                       tr(\Phi_{\alpha } \Phi_{\alpha }^{\dag }
              \wt{\Phi}_{\beta } \wt{\Phi}_{\beta}^{\dag }) \r\} \label{v1}
\eearr
The combined requirement of hermiticity and $U(1)_H \otimes U(1)_{PQ}$
 symmetry forces
all the parameters of $V_1(\Phi)$ to be real \cite{fnote1}. 
As a consequence, $CP$ appears as a symmetry of
$V_1(\Phi)$ although this was not imposed. One could choose a $CP$ conserving 
minimum for a suitable range of parameters :
\be
\langle \Phi _{\alpha}\rangle \equiv \left[ \barr{cc} \kappa_{\alpha u} & 0 \\
                                           0 & \kappa_{\alpha d}   
                                 \earr \right] \label{vev}
\ee                
where $\kappa_{{\alpha} u}$ and $\kappa_{{\alpha} d}$ are real.
 Eqs.(\ref{gamma})  
and (\ref{vev}) imply the following quark mass matrices:
\be
M_{u,d} = \left[ \barr{ccc} 0 & a\kappa_{1u,d} & 0 \\
                        a^*\kappa_{1u,d} & 0 & b\kappa_{2u,d} \\ 
                        0 & b^*\kappa_{2u,d} & c\kappa_{3u,d}  \earr \right] 
                                                               \label{qm} 
\ee
Note that the $M_u$ and $M_d$ allow for general up and down quark 
masses in spite of the
correlated structures. However because of this correlation,
 $M_u$ and $M_d$ can be 
simultaneously made real with a diagonal phase matrix $P$ :
\bearr
\widehat{M}_{u,d}  \equiv  P M_{u,d} P^{\dag}  
             & = & \left[\barr{ccc} 0 & |a|\kappa_{1u,d} & 0 \\
                         |a|\kappa_{1u,d} & 0 & |b|\kappa_{2u,d} \\
             0 & |b|\kappa_{2u,d} & |c|\kappa_{3u,d}  \earr \right] 
                                              \label{qmhat}   
\eearr
Phases in $P$ can be easily related to that in $a$ and $b$.
 $\widehat{M}_{u,d}$ 
are diagonalised by orthogonal matrices
$$ O_{u,d} \widehat{M}_{u,d} O^T _{u,d} = \h{diag}(m_{iu,d}) $$
Let us now discuss the $CP$ properties of the model.
Because of the fact that both $M_u$ and $M_d$ can be made real by the same 
phase matrix
$P$, the Kobayashi Maskawa matrices in the left as well as the right handed 
sectors are real. The reality of $\kappa _{\alpha u,d}$ also imply 
that the $W_L-W_R$ 
mixing is real. Hence gauge interactions are $CP$ conserving.
Moreover the matrix $P$ appearing in eq.(\ref{qmhat}) in fact make the 
individual Yukawa couplings real, i.e. 
\be P\Gamma _\alpha P^{\dag} = |\Gamma _\alpha |  \label{yukreal} \ee
for every $\alpha$. This has the consequence that the couplings of the 
neutral and charged Higgses
to the mass eigenstates  of quarks also become real. As a result, the Higgs
interactions would also conserve $CP$ as long as mixing among the 
Higgs fields is
$CP$ conserving. This is assured by the $CP$ invariance of
$V_1(\Phi)$ and reality of $\langle \Phi_{\alpha} \rangle $. 
It follows from the
above arguments that the model presented so far is in fact 
$CP$ conserving although
one did not impose it anywhere. 

We have not yet introduced fields needed to break 
$SU(2)_R \otimes U(1)_{PQ} \otimes
U(1)_H$. This can be done without spoiling the automatic $CP$
 invariance obtained
above.
As a concrete example let us introduce the conventional 
\cite{lr} $SU(2)$ triplet Higgses 
$\Delta _{L,R}$ with zero $U(1)_H$ and $U(1)_{PQ}$ charges. 
The breaking of the $PQ$ symmetry by $\langle \Phi
_{\alpha}\rangle $ generates a weak scale axion. 
We need to introduce a 
$G_{L,R} = SU(2)_L \otimes SU(2)_R \otimes U(1)_{B-L} $ singlet 
$ \sigma $ in order to make this axion invisible \cite{dine}.
  $ \sigma $ is taken to
transform under $PQ$ symmetry as $\sigma \rightarrow
e^{-i\beta} \sigma $ and remains invariant under $U(1)_H$. Finally,
 we introduce a
$G_{L,R}$ singlet field $\eta_H$ with $U(1)_H$ charge -2 and
transforming under the $PQ$ symmetry as $\eta_H \rightarrow
e^{-2i\beta} \eta_H $. The most general Higgs potential involving these
fields and their couplings to $\Phi $ fields can be written as:  
\bearr
V_2 & = &   \mu_{22} tr(\Phi_2 \widetilde{\Phi}_2 ^{\dag} ) \eta_H ^*  
         +\delta_{12} tr(\Phi_1 \widetilde{\Phi }_2 ^{\dag} ) \sigma^{*2}
 + \h{H.c} \non \\
   &   & + V(\Delta ) + V(\Delta \h{-}\Phi) +
        V(\eta_H \h{-}\sigma \h{-}\Delta \h{-}\Phi) \non 
\eearr 
For brevity, we do not display the parts $V(\Delta )$,
 $V(\Delta \h{-}\Phi)$ and
$V(\eta_H \h{-}\sigma \h{-}\Delta \h{-} \Phi )$ but mention that they 
contain only real couplings
\cite{dutta}. The only complex couplings possible are  $\mu_{22}$ and
$\delta_{12}$. But their phases can be absorbed into redefining 
$\sigma$ and $\eta_H$
without effecting reality of other parameters in $V_2$.   
Thus the above $V_2$ is automatically $CP$ conserving just like 
$V_1$ of eq.(\ref{v1}).
$V_1$ and $V_2$ together constitute the complete scalar potential 
of the model. 

We had imposed the discrete parity in the above analysis in order to obtain the
Fritzsch textures for $M_{u,d}$. But the automatic $CP$ invariance follows
 even in more
general situation without the discrete parity. In this case, $M_u$ and $M_d$
 are no
longer hermitian, but $U(1)_H$ symmetry still preserves texture zeroes 
appearing in
(13), (31), (11) and (22) elements of $M_u$ and $M_d$. It can be shown
 \cite{dutta}
that even in
this more general situation, the above argument goes through and one obtains 
automatic
$CP$ invariance. In contrast to the discrete parity, the left-right 
symmetry plays a
crucial role in giving the correlated structures for $M_{u,d}$ which 
lead to a $CP$
invariant theory.

Having presented a $CP$ invariant theory, we now discuss possible ways which 
lead
to small departures from exact $CP$ invariance. Obvious way is to enlarge the
Higgs sector in such a way that $CP$ gets violated in mixing among the Higgs
scalars. Alternative possibility is to assume that the horizontal symmetry gets
broken at a very high scale viz. grand unification scale. 
In this case \cite{ross} the $G$
invariant non-renormalizable couplings can induce sizable Yukawa coupling
at the low scales. This possibility is discussed by many authors 
\cite{ross} with a view of
understanding the textures of the fermion masses. In the present context, such
terms would also induce naturally small $CP$ violation. 
In fact the model presented
above allows the following general dim-5 terms resulting in fermion masses:   
 \be
-{\cal L}_{NR} = \frac{1}{M} \bar{q}_L {\Gamma}'_{\alpha} 
\widetilde{\Phi}_\alpha 
                 q_R \eta_H + \h{H.C.}     \label{nr}
\ee 
Here $M$ is some heavy mass scale which we take to be the Planck scale $M_P$.
 The
textures for ${\Gamma}'_\alpha $ are dictated by the $U(1)_H$ symmetry.
 The contribution
of ${\cal L}_{NR}$ to quark masses depends upon the parameter
 $\epsilon \equiv \frac{\langle
\eta_H \rangle }{M_P} $.

The $M_u$ and $M_d$ following from eqs. (\ref{qm}) and (\ref{nr}) can be 
written as
\cite{fnote2} :
\[
M_{u,d} = \left[\barr{ccc} 0 & a\kappa_{1u,d} & \epsilon \kappa_{3d,u}
 ({\Gamma}'_3)_{13} \\
                       a^*\kappa_{1u,d} &
 \epsilon \kappa_{3d,u} ({\Gamma}'_3)_{22} &
     b\kappa_{2u,d} + \epsilon \kappa_{2d,u} ({\Gamma}'_2)_{23} \\
\epsilon \kappa_{3d,u} ({\Gamma}'_3)^*_{13} & b^*\kappa_{2u,d} + \epsilon
         \kappa_{2d,u} ({\Gamma}'_2)^*_{23} & c\kappa_{3u,d} \earr \right]  
                                                                      \non
\]
The non-renormalizable contribution signified by $\epsilon$ works in a dual way
here. Firstly the presence of $\epsilon$ no longer makes it possible to rotate
away the phase from $M_{u,d}$ and hence from the KM matrix. Secondly it also
modifies the Fritzsch texture obtained in the above example. This is a welcome
feature in view of the fact that the Fritzsch ansatz is found to be
 inconsistent
\cite{nir} with the large top mass. The texture of $M_{u,d}$ obtained above
 retains the
successful predictions of the original ansatz and is also consistent
phenomenologically.

Note that the original Fritzsch ansatz implies that in the limit $\epsilon
\rightarrow 0$, 
\[
|a\kappa_{1u}|\sim \sqrt{m_u m_c}\; ;\;\; 
 |b\kappa_{2u}|\sim \sqrt{m_c m_t}\; ;\;\;
                               |c\kappa_{3u}|\sim m_t \; ;   
\]
\[
|a\kappa_{1d}|\sim \sqrt{m_d m_s}\; ;\;\; 
 |b\kappa_{2d}|\sim \sqrt{m_s m_b}\; ;\;\;
                                        |c\kappa_{3d}|\sim m_b     
\]
It follows therefore that $|\kappa_{2,3d}| \ll |\kappa_{2,3u}|$. 
Hence the presence of
$\epsilon$ terms alters the structure of $M_d$ more significantly 
than that of $M_u$. To
a good approximation \cite{dutta} one may take $M_u$ as in eq.(\ref{qmhat}) 
and $M_d$ as follows    
\be
     M_d \sim \left[ \barr{ccc} 0 & |a|\kappa_{1d} & 
                            \epsilon \kappa_{3u} \delta_1 e^{i\alpha}  \\
                             |a|\kappa_{1d} & \epsilon \kappa_{3u} \delta_2   
                                                & |b|\kappa_{2d} \\
                             \epsilon \kappa_{3u} \delta_1 e^{-i\alpha} & 
                                            |b|\kappa_{2d} & c\kappa_{3d}
                       \earr \r]                                \label{md}    
\ee
As before, we have redefined the quark fields and absorbed the phases 
of (12) and (23) 
elements. But this now leaves phases in terms involving $\epsilon$. 

Since the matrix diagonalising $M_u$ is completely fixed in terms of 
up-quark masses, we
can express $M_d$ of eq.(\ref{md}) in terms of the known parameters as 
\[
   M_d =  O_u^T{\cal K} \h{diag}(m_d,-m_s,m_b) {\cal K}^{\dag}O_u 
\]
 where ${\cal K}$ is the KM matrix in the Wolfenstein parameterization 
\cite{wolfenstein}. 
Comparing above
$M_d$ with the R.H.S of  eq.(\ref{md}) implies the successful relation
\[
\lambda = \sqrt{m_d \over m_s} - \sqrt{m_u \over m_c} 
\]
Moreover the other parameters also get fixed in terms of the masses 
and mixing angles.
Specifically,
\[
\begin{array}{l}
a\kappa_{1d}  \approx  -\sqrt{m_d m_s} \;\; ; \;\;
b\kappa_{2d}  \approx  - m_b \lambda^2 \l( A + {1 \over \lambda^2}
\sqrt{m_c \over m_t}\r) \;\; ; \;\; c\kappa_{3d}  \approx  m_b ; \\  
\epsilon \kappa_{3u} \delta_2  \approx  -m_s(1-\lambda^2) + m_b\l(
                        \lambda^2A + \sqrt{m_c \over m_t}\r)^2 ; \\
\epsilon \kappa_{3u} \delta_1 \cos \alpha  \approx  m_b A \lambda^3 \l
        ( \rho - {1\over \lambda}\sqrt{m_u \over m_c } \r) \;\; ; \;\; 
\epsilon \kappa_{3u} \delta_1 \sin \alpha  \approx  m_b A \lambda^3 \eta 
\end{array}
\]
where $A$, $\rho$ and $\eta$ are parameters in Wolfenstein matrix 
\cite{wolfenstein}.
The exact value of $\epsilon$ depends upon other parameters. 
If one chooses Yukawa couplings $c,  \delta_2 \sim O(1) $ then 
$ \epsilon \sim {m_s \over m_t}
\sim 10^{-3}$. Consistency then requires $ \delta_1 \sim 10^{-2} $ 
in this case.
 For $\epsilon \sim 10^{-3}$, the $ U(1)_H$ symmetry breaking scale is 
required to be of the order of $10^{16}$ GeV \cite{babu} 
if the scale of the non-renormalizable terms
is set by the Planck mass. 

In summary we have discussed one possible approach to understanding of 
small $CP$
violation in this paper. This is intimately linked to recent 
approaches which try to
understand the fermionic mass textures through higher 
dimensional terms generated by
flavor symmetry breaking at very large scale. 
This introduces in the low energy theory an effective small parameter 
controlling $CP$
violation. The horizontal symmetry cannot be directly probed in this case. 
Alternative
possibility not discussed here is to assume that 
horizontal symmetry is broken at low
$\sim $ TeV scale. In this case, $CP$ violation can be 
introduced \cite{dutta} through
enlargement in the Higgs sector. 
In either case, the conservation of $CP$ gets linked
intimately to the horizontal symmetry. 


\begin{thebibliography}{99}
\bibitem{km} M. Kobayashi and T. Maskawa, Prog. Theor. Phys.
 {\bf 49}, 652 (1973).
\bibitem{wali} See for example, 
A. R. Davidson and K. C. Wali, Phys. Rev. Lett. {\bf 48}, 11 (1982); 
         A.S. Joshipura, Z. Phys. {\bf C38}, 479 (1988).
\bibitem{ross} C. D. Foggartt and H. B. Nielsen, 
Nucl. Phys. {\bf B147}, 277 (1979);
               L. Ibanez and G. Ross, Phys. Lett. {\bf B332}, 100 (1994); 
               P. Binetruy and P. Ramond, Phys. Lett. {\bf B350}, 49 (1995); 
               V. Jain and R. Shrock, Phys. Lett. {\bf B352}, 83 (1995);
               Y. Nir, Phys. Lett. {\bf B354}, 107 (1995);
               B. Dudas, S. Pokorski and C. A. Savoy, Phys. Lett. {\bf B356},
 45 (1995);
               hep-ph/9509410.     
 
\bibitem{weinberg} S. Weinberg, Phys. Rev. Lett. {\bf 37}, 657 (1976);
                   G. C. Branco, Phys. Rev. Lett. {\bf 44}, 504 (1980).
\bibitem{fritzsch} H. Fritzsch, Nucl. Phys. {\bf B272}, 1519 (1986).
\bibitem{nir} F. Gilman and Y. Nir, Annu. Rev. Nucl. Part. Sci. 
{\bf 40}, 213 (1990);
              K. Kang, J. Flanz and E. Paschos, Z. Phys. {\bf C55}, 75 (1992).
\bibitem{lr}J. C. Pati and A. Salam, Phys. Rev.  {\bf D10}, 275 (1974);
   R. N. Mohapatra and J. C. Pati, Phys. Rev.  {\bf D11}, 566, 2558 (1975); 
   G. Senjanovic and R. N. Mohapatra, Phys. Rev. D {\bf 12}, 1502 (1975).
\bibitem{pq} R. D. Peccei and H. R. Quinn, Phys. Rev. Lett.
 {\bf 38}, 1440 (1977).
\bibitem{fnote1} The cyclic property of traces  has been used 
in arriving at this conclusion. 
\bibitem{dine} M. Dine, W. Fischler, M. Srednicki, Phys. Lett.
 {\bf B104}, 199 (1981).
\bibitem{dutta} Gautam Dutta and A. S. Joshipura, in preparation.
\bibitem{fnote2} We have assumed $M_u$, $M_d$ to be hermitian. 
This could follow  automatically from the discrete parity if
 $\eta_H$ also transforms non-trivially under it viz.
 $\eta_H \rightarrow \eta_H ^*$.  
\bibitem{wolfenstein} L. Wolfenstein, Phys. Rev. Lett. {\bf 51}, 1945 (1983).
\bibitem{babu} Note that $\langle \eta_H \rangle $ 
also breaks $PQ$ symmetry. The
 $\langle \eta_H \rangle \sim 10^{14} \h{GeV} $ falls 
outside the commonly accepted value
 for the $PQ$ scale namely $10^{10}-10^{12} \h{GeV} $.
 The upper bound on $PQ$ scale can
 however be relaxed, K. S. Babu, S. M. Barr and D. Seckel, 
Phys. Lett. {\bf B336 }, 213
(1994); G. Dvali, preprint IFUP-TH 21/95 (hep-ph/9505253). 
\end{thebibliography}
\end{document}